# COMPARATIVE WAKEFIELD ANALYSIS OF A FIRST PROTOTYPE OF A DDS STRUCTURE FOR CLIC MAIN LINAC

A. D'Elia[†*‡], R.M. Jones[†*], V.F. Khan[‡], A. Grudiev[‡], W. Wuensch[‡]
[†]School of Physics and Astronomy, The University of Manchester, Manchester, U.K.
[*] The Cockcroft Institute of Accelerator Science and Technology, Daresbury, U.K.
[‡] CERN, Geneva, Switzerland.

*Abstract*

A Damped Detuned Structure (DDS) for CLIC main linac has been proposed as an alternative to the present baseline design which is based on heavy damping. A first prototype, CLIC_DDS_A, for high power tests has been already designed and is under construction. It is also foreseen to design a further prototype, CLIC_DDS_B, to test both the wakefield suppression and high power performances. Wakefield calculations for DDS are, in the early design stage, based on single infinitely periodic cells. Though cell-to-cell interaction is taken into account to calculate the wakefields, it is important to study full structure properties using computational tools. In particular this is fundamental for defining the input parameters for the HOM coupler that is crucial for the performances of DDS. In the following a full analysis of wakefields and impedances based on simulations conducted with finite difference based electromagnetic computer code GdfidL will be presented.

## INTRODUCTION

The present CLIC baseline foresees a normal conducting structure supplying an average accelerating gradient of 100MV/m with an operating frequency of 11.994GHz [1], [2]. The RF design must minimize the electrical breakdown and ensure an efficient suppression of the beam-excited wakefields.

The first requirement can be achieved by changing the shape of the cavity walls which are designed to be elliptical. This ensures the pulsed temperature rise is minimised by re-distributing the surface fields [3], [4], [5]. A test structure known as CLIC_DDS_A has been designed and is presently under fabrication, under the supervision of KEK, to study the fundamental mode properties at high power operation (71MW peak input power) [5], [6], [7]. The fabrication of the structure is expected to be finished by the last quarter of 2011.

The complete method for damping the wakefields entails detuning the cell frequencies of the first dipole band in an error function fashion, by tapering down the irises along the structure. This spread prevents the wake from adding in phase. A moderate bandwidth of the lowest dipole frequencies was chosen so as to suppress the beam excited wakefields within an allowable limit for a revised inter bunch spacing of 8 rf cycles (0.67 ns) [5]. Due to the limited number of cells in the structure the wakefield recoheres. The recoherence can be suppressed by providing a moderate coupling ($Q$~1000) to four attached manifolds running parallel to the beam axis. Interleaving neighboring structure frequencies necessary to enhance wakefield suppression. A fully interleaved version (8-fold interleaved) of CLIC_DDS_A suppresses the wakefields within the beam dynamics level [4], [5]. This design is a proposed as a possible alternative to the present CLIC baseline which relies on heavy damping ($Q$~10 [8]) through strongly coupled waveguides attached to each cell.

The study of the wakefields in DDS requires a circuit model and Spectral Function method [9] of analysis. In this model the interaction between the cells is considered and its validity has been also demonstrated during NLC [9], [10]. However, due to the complexity of the engineering design, a cross check with computational tools is also needed. In the following a full comparison between GdfidL [11] and circuit model results will be reported.

## WAKEFIELD SIMULATIONS

A 3D CAD cross section cut-view of CLIC_DDS_A is shown in Fig. 1.

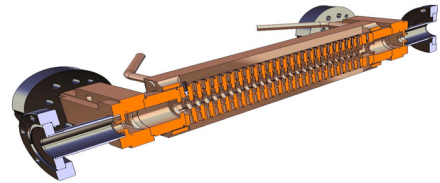

Figure 1: CLIC_DDS_A 3D cut-view.

The structure consists of 24 regular cells plus 2 cells at both ends to match the structure with a mode launcher. The manifold radius is constant along the structure and no HOM couplers are foreseen as the purpose at this stage is to test its ability to sustain high powers.

In order to properly suppress the wakefield eight of these structures with interleaved dipole frequencies are needed. However, for the sake of comparison of wakefields we simulated the wakefield in a single CLIC_DDS_A structure with GdfidL and compare the results with the prediction of the circuit model and spectral function method. Modeling eight interleaved structures with GdfidL, or a similar computer code, will be prohibitively expensive from a computational point of view. This is the reason for the necessity of the circuit model. Confidence in the ability to predict wakefield with this model is provided by this limited structure comparison. Figure 2 illustrates the 3D model used in GdfidL to represent 24 cell structure of length ~0.2 m.

The mesh spacing, in cartesian coordinates is 100 μm and a bunch of charge 1pC and length of $\sigma_z$~500 μm has been considered.

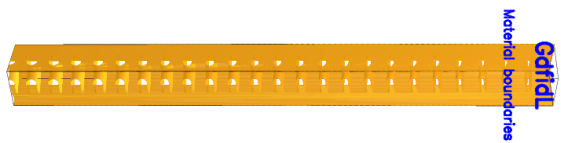

Figure 2: GdfidL mesh model: a quarter of the structure is considered with magnetic and electric boundary condition on cutting planes; the structure is terminated downstream and upstream with perfectly matched layers.

The transverse wakefield, simulated with GdfidL is illustrated in Fig. 3, together with the impedance, obtained from the Fourier Transform thereof.

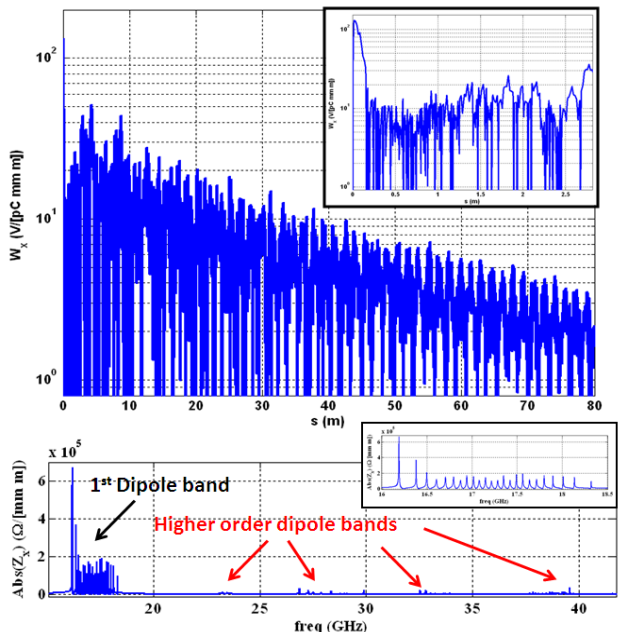

Figure 3: Result of GdfidL simulation. Top: transverse wakefield, inset the earliest 3m. Bottom: absolute value of the transverse impedance dominated by 1st dipole band (inset).

## SPECTRAL FUNCTION METHOD COMPARED WITH GDFIDL

The circuit model and spectral function method have been used to calculate the envelope of the transverse wakefield [4]. This allows a rapid determination of the wakefield and enables the extent of fabrication tolerances to be effortlessly taken in to account [10]. Here we compare the model with results obtained from the GdfidL code. The latter code requires considerable computational resources in terms of time and memory in order to simulate the full 24 cell structure. Indeed for the parameters used, up to 80 m of wakefield, almost one week is required to perform the full simulation.

A Fourier transform is performed to obtain the impedance which necessitates 0-padding and re-interpolation. The resonances in the impedance are subsequently fitted with a Lorentzian function to obtain the $Q$s, frequencies and kick factors to be extracted for each mode. The results of this procedure, compared with those of the spectral function method are illustrated in Fig. 4.

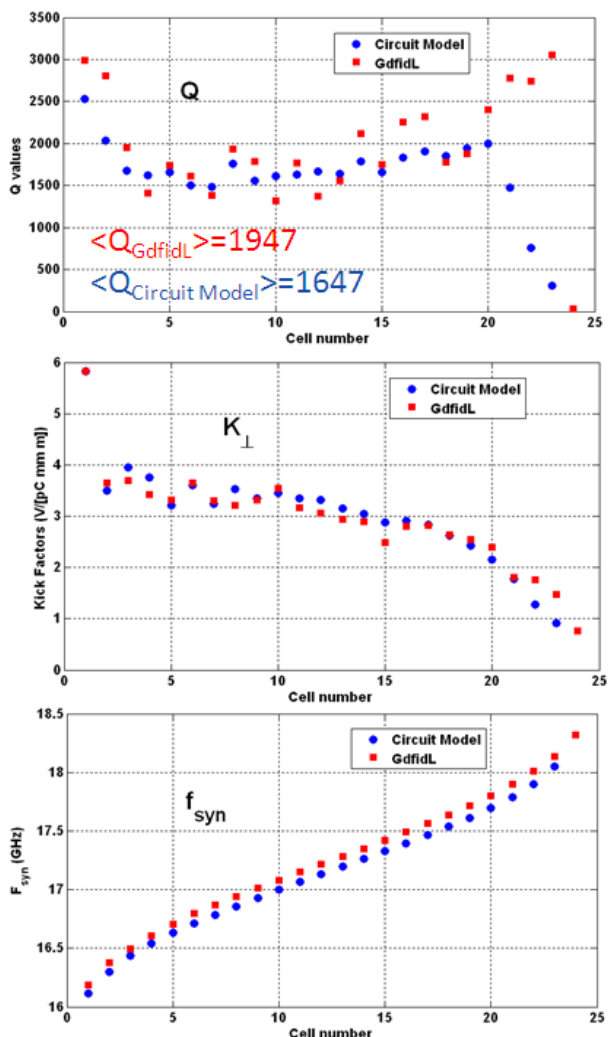

Figure 4: Dipolar Q's, Kick Factors and Fsyn's of the first dipolar band from GdfidL (in red) and from Circuit Model (in blue).

The agreement is reasonable in all cases apart from a systematic frequency shift of 90 MHz and significantly differing $Q$s. However, Gdfidl is not able to directly compute the effect of Cu wall losses and this accounts for the major discrepancies in Qs. Furthermore, the fundamental mode coupler loads damps the last cell with a Q = 35 and this is included in the circuit model only.

The wakefields can be expressed as a sum over the coupled modes [9]

$$W_\perp(s) = 2\left|\sum_{n=1}^{N} K_{\perp n}\, Exp\left\{i\frac{\omega_n s}{c}\left(1+\frac{i}{2Q_n}\right)\right\}\right| \quad (1)$$

where $K_{\perp n}$, $\omega_n/2\pi$ and $Q_n$ are the kick factor, synchronous frequency and Q of the n-th mode of an N-cell structure.

We use this approximate representation of the wakefield to include Cu wall losses into the GdfidL simulations. However, prior to inserting losses it is interesting to isolate the dipole modes from the overall wakefield produced from GdfidL simulations. This is achieved by using Eq. (1) without losses from the parameters obtained in the Lorentzian fits of the dipole impedance. The results of this study are presented in Fig. 5. Simulations using several computer 2D codes [12] have indicated that the dominant component of the wakefield is located in the first dipole band and this is compared in Fig. 5.

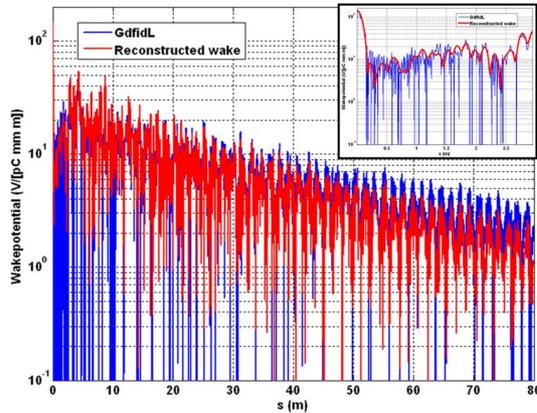

Figure 5: GdfidL multipole wake in blue compared with dipole component of GdfidL wake according to Eq. (1); inset shows a particular of the earliest 3m.

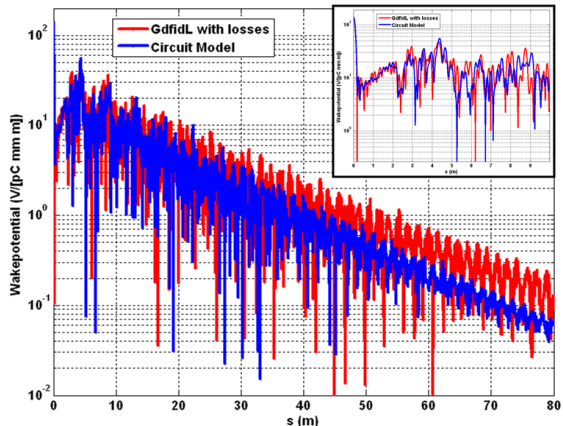

Figure 6: Reconstructed GdfidL wake including Cu losses compared with the circuit model wake; the earliest 10m is shown inset.

Adding losses to all modes, with a $Q \sim 6000$ (from single cell simulations in HFSS [13]) apart from the last cell where Q=35, is illustrated in Fig. 6. The wake of both the circuit model and GdfidL compare well especially up to 10 m. The circuit model includes a final cell with a Q of 35 and this will be shared amongst adjacent cells to form the modal Q. However, in the artificial reconstruction of the wake with GdfidL, the modal Q is assigned a value of 35. This is clearly an approximation and will not properly represent the coupling. It is difficult to improve on this approximate interpretation of modal Qs as inferred from their cell counterparts. This accounts for some of the discrepancies in the Qs of the resonances in the impedances. The impedance of the dipole modes (including Cu losses) obtained from GdfidL is shown in Fig. 7 and compared with the spectral function method. Again, reasonable agreement between different techniques is obtained.

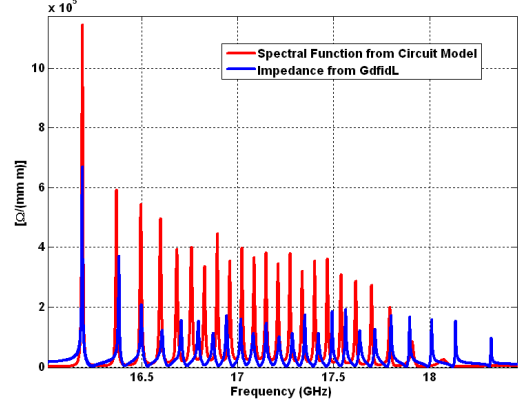

Figure 7: Comparison between Spectral Function in red and transverse impedance from GdfidL in blue (frequency shifted systematically by 90 MHz).

## FINAL REMARKS

Reasonable agreement has been obtained between different simulation methods. Additional work is in progress on designing suitable higher order mode coupler for these structures.

## ACKNOWLEDGEMENTS

Research leading to these results has received funding from European commission under the FP7 research infrastructure grant no. 227579.